\documentclass[dvips]{acta}
\usepackage{supertabular,lscape,epsfig}
\usepackage{amssymb}
\usepackage{amsmath}
\DeclareSymbolFont{ppa}{OT1}{ppl}{m}{it}
\DeclareMathSymbol{\vv}{\mathalpha}{ppa}{'166}

\SetPages{293}{312}

\SetVol{58}{2008}

\begin{document}

\newcommand{\TabCapp}[2]{\begin{center}\parbox[t]{#1}{\centerline{
  \small {\spaceskip 2pt plus 1pt minus 1pt T a b l e}
  \refstepcounter{table}\thetable}
  \vskip2mm
  \centerline{\footnotesize #2}}
  \vskip3mm
\end{center}}

\newcommand{\TTabCap}[3]{\begin{center}\parbox[t]{#1}{\centerline{
  \small {\spaceskip 2pt plus 1pt minus 1pt T a b l e}
  \refstepcounter{table}\thetable}
  \vskip2mm
  \centerline{\footnotesize #2}
  \centerline{\footnotesize #3}}
  \vskip1mm
\end{center}}

\newcommand{\MakeTableSepp}[4]{\begin{table}[p]\TabCapp{#2}{#3}
  \begin{center} \TableFont \begin{tabular}{#1} #4
  \end{tabular}\end{center}\end{table}}

\newcommand{\MakeTableee}[4]{\begin{table}[htb]\TabCapp{#2}{#3}
  \begin{center} \TableFont \begin{tabular}{#1} #4
  \end{tabular}\end{center}\end{table}}

\newcommand{\MakeTablee}[5]{\begin{table}[htb]\TTabCap{#2}{#3}{#4}
  \begin{center} \TableFont \begin{tabular}{#1} #5
  \end{tabular}\end{center}\end{table}}

\newfont{\bb}{ptmbi8t at 12pt}
\newfont{\bbb}{cmbxti10}
\newfont{\bbbb}{cmbxti10 at 9pt}
\newcommand{\uprule}{\rule{0pt}{2.5ex}}
\newcommand{\douprule}{\rule[-2ex]{0pt}{4.5ex}}
\newcommand{\dorule}{\rule[-2ex]{0pt}{2ex}}
\def\thefootnote{\fnsymbol{footnote}}

\hyphenation{Ce-phe-ids eclip-ses chan-ges me-thod}

\begin{Titlepage}
\Title{The Optical Gravitational Lensing Experiment.\\
The OGLE-III Catalog of Variable Stars.\\
II. Type II Cepheids and Anomalous Cepheids\\
in the Large Magellanic Cloud\footnote{Based on
observations obtained with the 1.3-m Warsaw telescope at the Las Campanas
Observatory of the Carnegie Institution of Washington.}}
\Author{I.~~S~o~s~z~y~\'n~s~k~i$^1$,~~
A.~~U~d~a~l~s~k~i$^1$,~~
M.\,K.~~S~z~y~m~a~\'n~s~k~i$^1$,\\
M.~~K~u~b~i~a~k$^1$,~~
G.~~P~i~e~t~r~z~y~\'n~s~k~i$^{1,2}$,~~
\L.~~W~y~r~z~y~k~o~w~s~k~i$^3$,\\
O.~~S~z~e~w~c~z~y~k$^2$,
~~K.~~U~l~a~c~z~y~k$^1$~~
and~~R.~~P~o~l~e~s~k~i$^1$}
{$^1$Warsaw University Observatory, Al.~Ujazdowskie~4, 00-478~Warszawa, Poland\\
e-mail:
(soszynsk,udalski,msz,mk,pietrzyn,wyrzykow,szewczyk,kulaczyk,rpoleski)
@astrouw.edu.pl\\
$^2$ Universidad de Concepci{\'o}n, Departamento de Fisica, Casilla 160--C,
Concepci{\'o}n, Chile\\
e-mail: szewczyk@astro-udec.cl\\
$^3$ Institute of Astronomy, University of
Cambridge, Madingley Road, Cambridge CB3 0HA, UK\\
e-mail: wyrzykow@ast.cam.ac.uk}
\Received{November 22, 2008}
\end{Titlepage}
\Abstract{In the second part of the OGLE-III Catalog of Variable Stars
(OIII-CVS) we present 197 type~II Cepheids and 83 anomalous Cepheids in the
Large Magellanic Cloud (LMC). The sample of type~II Cepheids consists of 64
BL~Her stars, 96 W~Vir stars and 37 RV~Tau stars. Anomalous Cepheids are
divided into 62 fundamental-mode and 21 first-overtone pulsators. These are
the largest samples of such types of variable stars detected anywhere
outside the Galaxy.

We present the period--luminosity and color--magnitude diagrams of stars in
the sample. If the boundary period between BL~Her and W~Vir stars is
adopted at 4~days, both groups differ significantly in $(V-I)$ colors. We
identify a group of 16 peculiar W~Vir stars with different appearance of
the light curves, brighter and bluer than ordinary stars of that type. Four
of these peculiar W~Vir stars show additional eclipsing modulation
superimposed on the pulsation light curves. Four other stars of that type
show long-period secondary variations which may be ellipsoidal
modulations. It suggests that peculiar W~Vir subgroup may be related to
binarity. In total, we identified seven type~II Cepheids simultaneously
exhibiting eclipsing variations which is a very large fraction compared to
classical Cepheids in the LMC. We discuss diagrams showing Fourier
parameters of the light curve decomposition against periods. Three sharp
features interpreted as an effect of resonances between radial modes are
detectable in these diagrams for type~II Cepheids.}{Cepheids -- Stars:
oscillations -- Stars: Population II -- Magellanic Clouds}

\Section{Introduction}
\vspace*{12pt}
Until the mid of 20th century astronomers were not aware that there exist
two different types of Cepheid variables. Only after the famous distance
scale revision done by Baade (1952), it became clear that Population I and
II Cepheids follow different period--luminosity (PL) relations. On the
average type~II Cepheids are about 1.5~mag fainter than classical Cepheids
of the same periods.

Type II Cepheids (also referred to as Population II Cepheids, although
Feast \etal 2008 noted that Population II Cepheids are a subset of the
type~II Cepheids) consist of three subclasses in different evolutionary
stages: BL~Her, W~Vir and RV~Tau stars. The shortest-period variables --
BL~Her stars -- are evolving from the horizontal branch toward the
asymptotic giant branch (AGB). This group is sometimes called AHB stars
(``above horizontal branch'', Strom \etal 1970, Kraft 1972). The W~Vir
stars cross the instability strip during their blue-loop excursions from
the AGB during helium-shell flashes. The brightest stars in the Population
II instability strip are RV~Tau stars, which are at the stage of leaving
the AGB on the way to the white dwarf domain. A defining characteristic of
the RV~Tau stars are alternating deep and shallow minima of their light
curves.

Anomalous Cepheids (sometimes called BL~Boo stars) are metal-poor stars
which spread between type I and type~II Cepheids in the PL diagram. The
pulsation masses of anomalous Cepheids are estimated to be about 1.5~\MS
(\eg Wallerstein and Cox 1984, Bono \etal 1997). These stars are believed
to be formed through the merging of a binary star system and may be related
to blue stragglers. The other possibility is that anomalous Cepheids are
much younger than other metal-poor stars in their systems. Most of the
known anomalous Cepheids were found in nearby dwarf spheroidal galaxies,
only a few such stars were identified in globular clusters (Zinn and Dahn
1976, Kaluzny \etal 1997).

The first systematic search for type~II Cepheids in the Large Magellanic
Cloud (LMC) was performed by Hodge and Wright (1963), who surveyed the
vicinities of the LMC old clusters. However, as subsequent studies showed,
all of the 11 periodic variables detected by Hodge and Wright (1963) turn
out to be classical Cepheids. The last object from their list -- HV~12904
-- was reclassified as a classical Cepheid (OGLE-LMC-CEP-0318) in the
previous part of this catalog (Soszy{\'n}ski \etal 2008, hereafter
Paper~I). The first two reliable type~II Cepheids in the LMC (HV~2351 and
HV~5690) were identified by Hodge and Wright (1969). Then, the significant
sample of 17 type~II Cepheids in the LMC was included in the catalog of
variable stars of Payne-Gaposchkin (1971). The discovery of the first
RV~Tau star in the LMC (HV~13066) was reported by Wright and Hodge (1971).

The next breakthrough in this field came with the large microlensing
surveys: MACHO and OGLE. Alcock \etal (1998) presented a list of 33 W~Vir
and RV~Tau stars in the LMC. In the PL diagram the RV~Tau stars appeared to
be a direct extension of other type~II Cepheids to longer periods. The
OGLE-II catalog of Cepheids in the LMC (Udalski \etal 1999) contained 35
type~II Cepheids which were compared with the bulge variables of that type
by Kubiak and Udalski (2003).

Anomalous Cepheids in the LMC have been practically unknown to date. The
extragalactic part of the General Catalogue of Variable Stars (GCVS,
Artyukhina \etal 1995) lists no such objects in the LMC. Di~Fabrizio \etal
(2005) reported the discovery of 4 anomalous Cepheids in this galaxy, but
we prefer to classify these stars as RR~Lyr variables (see Section~3.2).

In this paper we describe the second part of the OGLE-III Catalog of
Variable Stars (OIII-CVS) containing 197 type~II and 83 anomalous
Cepheids in the LMC. These are the largest samples of such objects ever
found in any extragalactic system. We discuss the classification
criteria, present a preliminary statistical analysis of these Cepheids
and show objects worth particular interest.

\Section{Observations and Data Reduction}
OGLE-III observations were carried out with the 1.3~m Warsaw telescope at
Las Campanas Observatory, Chile, operated by the Carnegie Institution of
Washington. The telescope is equipped with a large field CCD mosaic camera,
consisting of eight $2048\times4096$ pixel SITe ST002A detectors. The pixel
size of each of the detectors is 15~$\mu$m giving the 0.26~arcsec/pixel
scale and field of view $35\times35.5$~arcmins. The details of the system
can be found in Udalski (2003).

The observations span 2400 days from July 2001 to March 2008. A total sky
coverage in the LMC was 39.7 square degrees. The photometry of stars in the
central parts of the LMC is supplemented by the OGLE-II photometry
collected between 1997 and 2000. Photometry is in the standard {\it VI}
bands with about 90\% of observations ($\approx400$ points per star) in the
{\it I}-band. The exposure time was 180~sec and 225~sec in the {\it I-} and
{\it V}-bands, respectively.

The photometry was obtained using the Difference Image Analysis (DIA)
method -- image subtraction algorithm developed by Alard and Lupton (1998)
and Alard (2000), and implemented by Wo{\'z}niak (2000). Udalski \etal
(2008) gives full details of the data reduction techniques. The
uncertainties of the photometry were corrected using methods developed by
J.~Skowron and described in Paper~I.

\begin{figure}[p]
\centerline{\includegraphics[width=13.5cm, bb=35 55 545 745]{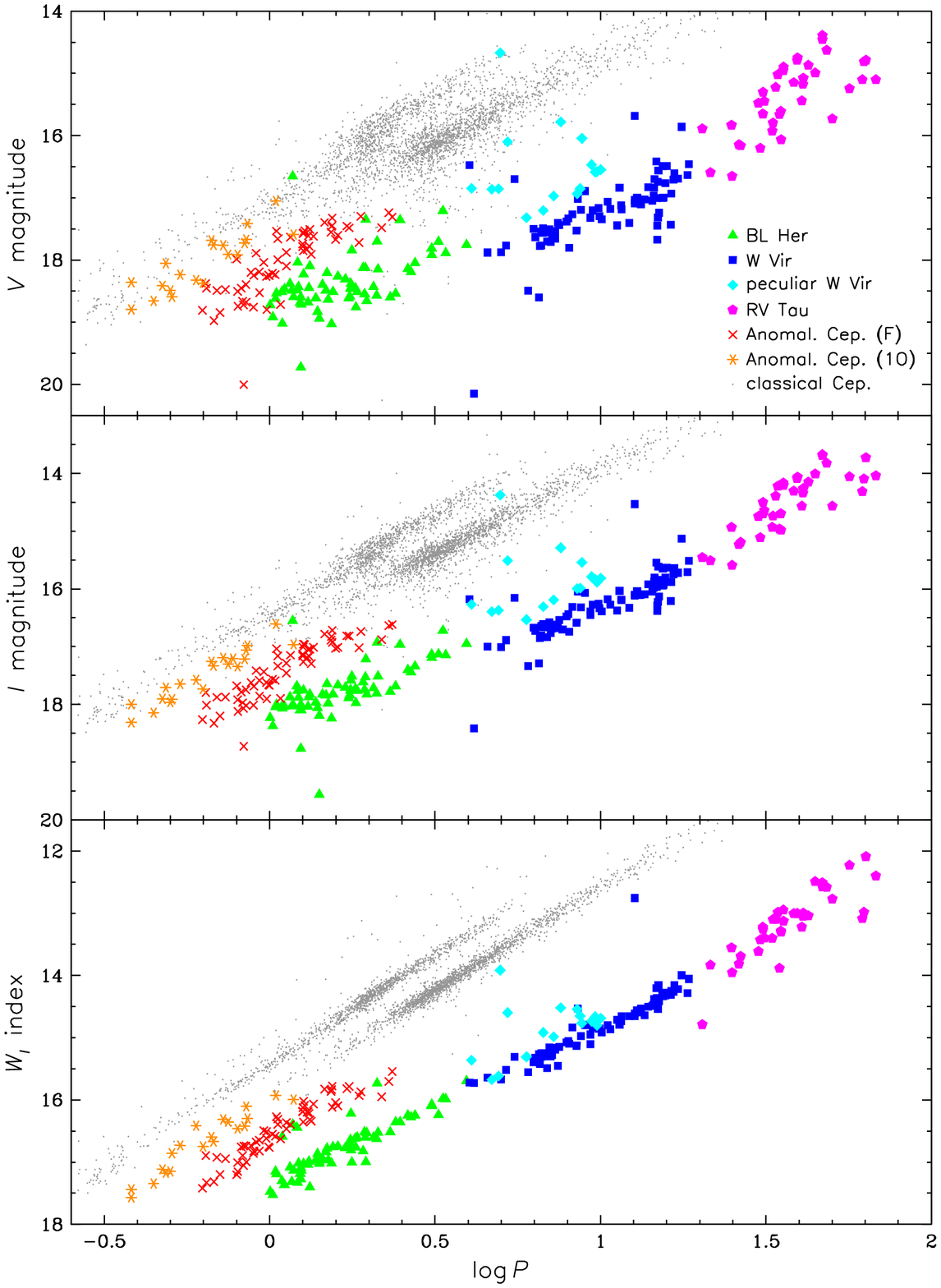}}
\FigCap{Period--luminosity diagrams for Cepheids in the LMC. Green, blue,
cyan and magenta symbols show type~II Cepheids, red and orange -- anomalous
Cepheids, grey points represent classical Cepheids from Paper~I.}
\end{figure}

\Section{Selection and Classification of Variables}
\Subsection{Type II Cepheids}
Type II Cepheids in the LMC were selected using the same methods as for
classical Cepheids presented in the previous part of the OIII-CVS.
The selection process started with a massive period search
conducted at the Interdisciplinary Centre for Mathematical and
Computational Modelling of Warsaw University (ICM). The calculations
were performed for all stars observed by OGLE in the LMC, in total 32
million objects, using program {\sc FNPeaks} written by
Z.~Ko{\l}aczkowski. Then, we selected a subsample of periodic variables
located in the PL diagrams inside a wide strip covering classical and
type~II Cepheids. We used PL diagrams plotted for the visual
luminosities ({\it V}, {\it I}, Wesenheit index $W_I$), as well as the
near-infrared magnitudes from the 2MASS Point Source Catalog (Cutri
\etal 2003).

The light curves chosen in such a way were visually inspected and divided
into three groups -- pulsating-like, eclipsing-like and other variables. In
the next phase the pulsating variables were tentatively categorized into
classical Cepheids, type~II Cepheids, anomalous Cepheids, RR~Lyr stars and
$\delta$~Sct stars. This classification was based on the position of the
stars in the PL diagrams. The objects located outside the instability strip
in the color--magnitude (CM) diagram were removed from the sample, however
it was checked if these objects are not highly reddened by extinction or
have erroneous photometry. All the stars lying close to the boundaries
between the regions in the PL diagrams, or with unusual positions in the
diagrams showing parameters of the Fourier decomposition \vs periods, were
treated in particular way. The classification was changed for some stars,
because it was recognized that these objects are blended.

In the second stage of the type~II Cepheids selection we repeated the
visual inspection of the periodic light curves collected in our
database. This time we chose only those stars which were located in the PL
and CM diagrams close to the regions occupied by type~II Cepheids
identified so far. In this way we changed the previous classification for
several variables, for example, we noticed some RV~Tau stars with light
curves almost indistinguishably similar to eclipsing variables. Only a
careful analysis of these light curves revealed a small asymmetry which
indicates that these stars are pulsating variables. One should, however, be
aware that our final classification may be uncertain in some cases. In
several stars of exceptionally ``noisy'' light curves we were unable to
reliably distinguish between eclipsing/ellipsoidal stars and pulsating
variables. We flagged these objects as uncertain in our catalog.

The final sample of type~II Cepheids contains 197 objects. The PL diagrams
of these variables in $V$, $I$ and extinction insensitive Wesenheit index
$W_I=I-1.55(V-I)$ are plotted in Fig.~1. Significant difference in the
scatter of the relations in $V$ and $W_I$ domains suggests that a
remarkable number of the stars are affected by reddening. Several outliers
visible in the $\log{P}{-}W_I$ diagram are possibly blended stars
(including ``eclipsing'' Cepheids -- see Section~5) but some of them can be
atypical Cepheids. In Section~4 we describe a group of peculiar W~Vir stars
(marked with cyan symbols in Fig.~1), brighter than ordinary stars of this
type.

The RV~Tau stars are plotted with ``single'' periods, \ie defined as the
intervals between successive minima. As one can notice, the PL relation are
not linear for the whole range of periods covered by type II Cepheids, and
actually the PL relations should be fitted separately for BL~Her, W~Vir and
RV~Tau stars. This conclusion is especially valid for RV~Tau stars which
seem to be much brighter than would be expected from the extrapolated
relation fitted to shorter-period type~II Cepheids (\eg Demers and Harris
1974, Alcock \etal 1998).

The number density in the period distribution of type~II Cepheids peaks
around three values, corresponding to BL~Her, W~Vir and RV~Tau stars. We
separated these groups using the following criteria. As the limiting period
defining the boundary between RR Lyr and BL~Her stars we adopted
1~day. This is the most often used value (\eg Arp 1955, Payne-Gaposchkin
1956, Harris 1985, Clement \etal 2001), though other limiting periods were
also sometimes used, like $P=0.75$~days in Wallerstein and Cox (1984) or
$P=0.8$~days in Gautschy and Saio (1996). However, our choice of the
boundary period was not motivated only by tradition. We found that the
distribution of Population II pulsating stars has the minimum for
$P=1$~day. Moreover, we noticed a small discontinuity (drop from 5.5 to
4.5) for this period in the $\log{P}{-}\phi_{21}$ diagram (see Section~7).

\begin{figure}[t]
\centerline{\includegraphics[width=12.1cm, bb=35 275 565 745]{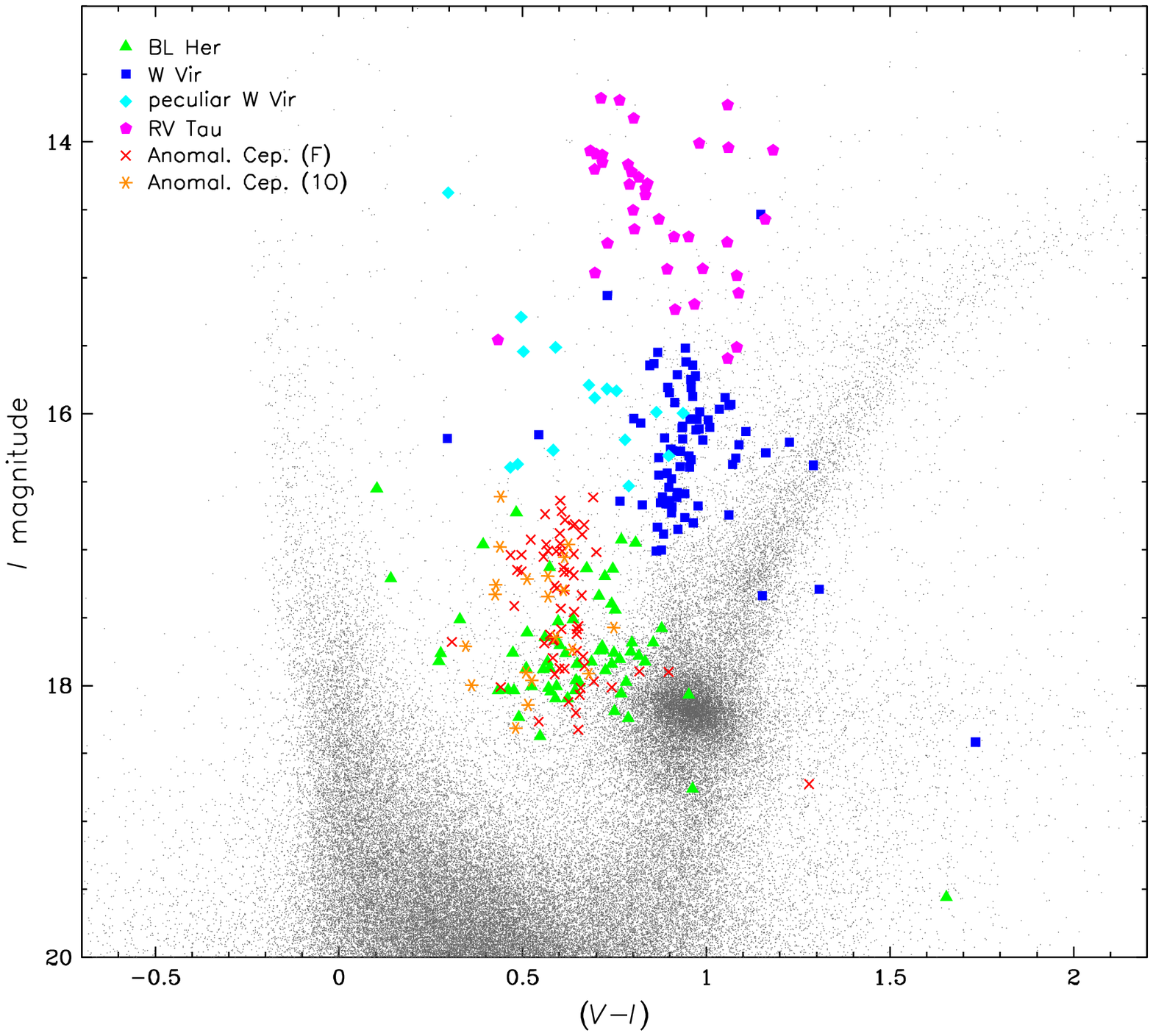}}
\vskip4pt
\FigCap{Color--magnitude diagram for type~II and anomalous Cepheids in the
LMC. The color symbols are the same as in Fig.~1. In the background all
stars from the subfield LMC100.1 are shown.}
\end{figure}

The transition between BL~Her and W~Vir stars is also not clearly
defined. The boundary periods provided in the literature range from 3 to
10~days. We decided to define the threshold between both subtypes of
type~II Cepheids at $P=4$~days, because around this value we noticed the
frequency minimum of the sample. Moreover, for $P=4$~days we found a clear
discontinuity in the CM diagram (Fig.~2) -- a~shift in the mean $(V-I)$
colors from about 0.6 to 0.9~mag.

The distinction between W~Vir and RV~Tau stars is somewhat ambiguous. It
has been known for years that there is an overlap in the properties of
W~Vir and RV~Tau stars. Sometimes stars classified as W~Vir variables do
show alternating depth of minima (Arp 1955). Besides, infrared excess
usually observed in RV~Tau variables is also detectable in some W~Vir stars
(Welch 1987, Laney 1991).
\begin{figure}[b]
\vspace{-3mm}
\centerline{\includegraphics[width=11.5cm]{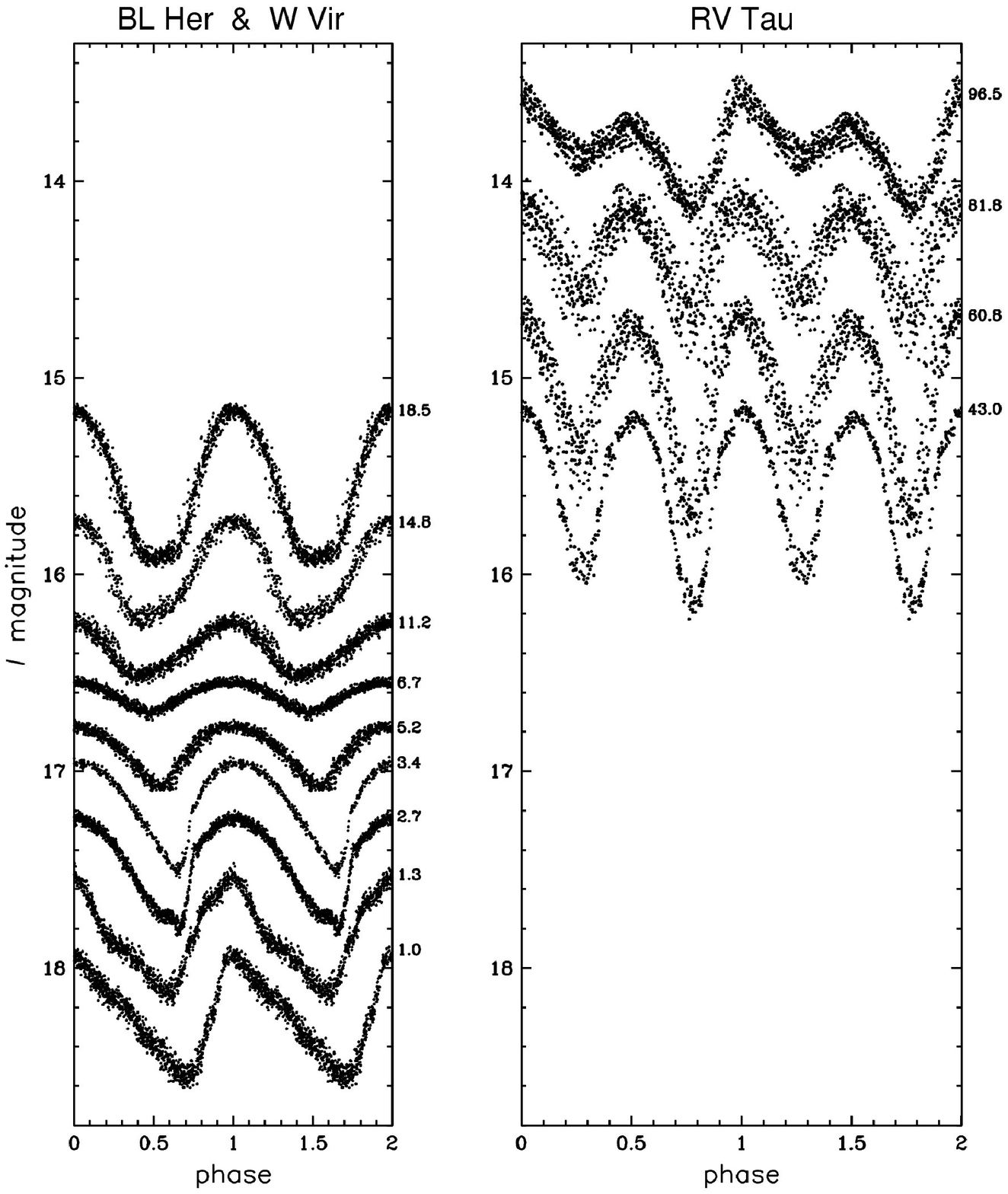}}
\vspace{-3mm}
\FigCap{Exemplary light curves of type~II Cepheids in the LMC. {\it Left 
panel} shows BL~Her and W~Vir stars, while {\it right panel} presents RV~Tau
stars phased with ``double'' (formal) periods. Small numbers at the right
side of each panel show the rounded periods in days of the light curves
presented in panels.}
\end{figure}

In our sample of type~II Cepheids all stars with periods longer than
20~days are considered to be RV~Tau stars. Again, we noticed a local
minimum in the period-frequency distribution around this period. The
vast majority of variables above this limit exhibit alternating deep and
shallow minima, or at least larger scatter of points in the minima, what
may be caused by switching deep and shallow minima observed in many
RV~Tau stars. It is important to mention that our criterion categorized
as a W~Vir star at least one star with distinct alternation of cycles 
(OGLE-LMC-T2CEP-002 with semi-period equal to 18.3~days) and several
stars with equal minima fell within the RV~Tau range of periods.

The longest (single) period of the RV~Tau star in our sample is equal to
68~days. Above this limit we found some semiregular variables with a
behavior similar to the RV~Tau stars, \ie alternating deep and shallow
minima. However, these stars were removed from our list, because their
colors were much redder than for typical Cepheids. These stars follow the
PL relations for semiregular variables (Soszy{\'n}ski \etal 2007) and will
be included in the sample of variable red giants which will be published in
a future paper.

Finally, we divided the sample into 64 BL~Her, 96 W~Vir and 37 RV~Tau
stars. Fig.~3 shows typical {\it I}-band light curves of type~II
Cepheids. Light curves of BL~Her and W~Vir stars are presented in the left
panel. The right panel shows four RV~Tau stars phased with the ``double''
(formal) period. Type~II Cepheids show a wide variety of light curve
shapes, although generally there is a clear progression of light curve
shapes with period. However, some diversity of the observed light curves is
expected, because of the great sensitivity of the light curve shapes on
stellar parameters (Moskalik and Buchler 1993). Fig.~3 does not contain
peculiar W~Vir stars detected in our data, exhibiting somewhat different
colors, magnitudes and light curve appearance than regular variables of
that type (see Section~4).

Note that, with exception of the two faintest stars, the considerable
scatter visible in the phased light curves is not caused by photometric
errors. For many stars from our sample this scatter is even more
significant which is a well-known feature of type~II Cepheids. Most of the
unstable light curves can be explained by a variable period.

\Subsection{Anomalous Cepheids}
After the manual selection of pulsating stars in our database and plotting
them in the PL diagram (Fig.~1), the sequence of fundamental-mode anomalous
Cepheids was clearly visible. This PL relation is located between classical
and type~II Cepheid strips, up to the periods of about 2.4~days. Thus, the
identification of anomalous Cepheids with periods exceeding 1~day was a
relatively easy task, also because these stars occurred to constitute quite
homogeneous class of variables. Light curves of several anomalous Cepheids
are shown in Fig.~4. As one can notice the fundamental-mode pulsators have
asymmetric light curves with rapid rise to maximum and slow decline. The
vast majority of these stars exhibit a small bump just before the rise to
maximum light.

\begin{figure}[t]
\vspace{-3mm}
\centerline{\includegraphics[width=10.5cm]{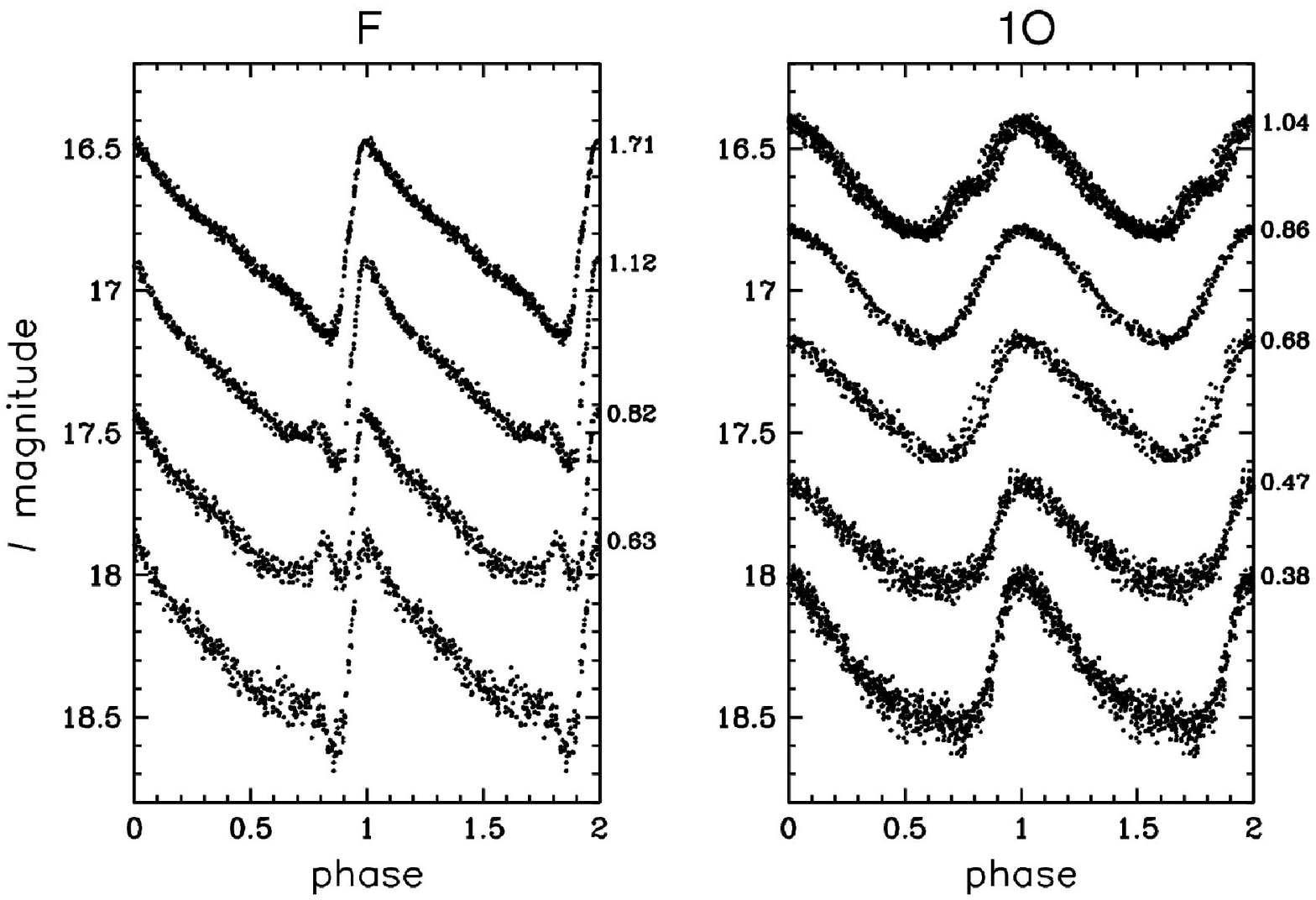}}
\FigCap{Exemplary light curves of anomalous Cepheids in the LMC. {\it Left 
panel} shows fundamental-mode pulsators, while {\it right panel} presents
first-overtone stars. Small numbers at the right side of each panel show
the rounded periods in days of the light curves presented in panels.}
\end{figure}

The situation with fundamental-mode anomalous Cepheids with periods below
1~day was more complicated. Their light curves practically have the same
morphology as fundamental-mode RR~Lyr (RRab) stars, so, in general, it is
impossible to photometrically distinguish between the LMC short-period
anomalous Cepheids and RR~Lyr stars located somewhat closer to
us. Moreover, we detected a considerable number of blended RR~Lyr stars
with brightness increased by unresolved stars. These objects should have
smaller amplitudes in the magnitude scale, but it is not always
unambiguously detectable, because RR Lyr variables cover wide range of
typical amplitudes.

Our selection criteria of short-period anomalous Cepheids were based on the
features of these stars with $P>1$~day. We searched for objects brighter
than typical RR~Lyr stars and with light curves similar to those presented
in the left panel of Fig.~4. Since we have not observed Blazhko effect for
long-period anomalous Cepheids, we also excluded light curves with
considerable scatter which usually is caused by this phenomenon. We
emphasize that our procedure might not be perfect and some of the
short-period anomalous Cepheids could actually be RR~Lyr stars. 62
fundamental-mode anomalous Cepheids with periods between 0.62 and 2.35~days
passed our selection criteria.

The candidates for anomalous Cepheids pulsating in the first overtone were
selected on the basis of their position in the PL diagram and the
appearance of their light curves. Several such light curves are presented
in the right panel of Fig.~4. Generally, the light curves are smoother than
for the fundamental-mode pulsators, with rounded maxima and minima,
although some stars exhibit somewhat sharper maximum. The light curves
resemble two known Galactic anomalous Cepheids: BL~Boo (Zinn and Dahn 1976)
and XZ~Cet (Teays and Simon 1985), which are proved to be overtone
pulsators (\eg Zinn and King 1982, Szabados \etal 2007). The PL relation of
the overtone anomalous Cepheids falls on the extension to shorter periods
of the fundamental-mode classical Cepheids, but the light curves of both
groups are distinctly different. We found only two first-overtone anomalous
Cepheid with period longer than 1~day (OGLE-LMC-ACEP-015,
OGLE-LMC-ACEP-050), \ie in the range occupied by classical Cepheids. These
objects have been distinguished on the basis of their light curve
morphology. In total, we selected 21 candidates for overtone anomalous
Cepheids with periods ranging in 0.38--1.18~days.

It is remarkable that none of the 4 candidates for anomalous Cepheids in
the LMC detected by Di Fabrizio \etal (2005) passed our selection
criteria. Two of these stars have decreased amplitudes and are probably
blended RR~Lyr stars, one is a Blazhko RRab star with the mean {\it I}-band
magnitude equal to 18.4~mag (\ie in the LMC RR~Lyr range), and one is
probably an RRc star, although it is fainter than other stars of that type
($I=19.5$).

\Section{Peculiar W~Vir stars}
During the inspection of the light curves we noticed a group of W~Vir stars
with periods between about 6 and 10 days with somewhat different light
curve shapes than other stars of that type. The typical type~II Cepheid
with the period in this range shows a light curve with slow rise and more
rapid decline, or, at least, a symmetric light curve with broad flat
maximum. We noticed several Cepheids with the rising branch steeper than
declining one. In this paper we call these objects ``peculiar W~Vir
stars''. Fig.~5 compares light curves for the peculiar and ordinary W~Vir
stars at similar periods.

\begin{figure}[htb]
\vspace{-2mm}
\centerline{\includegraphics[width=13cm]{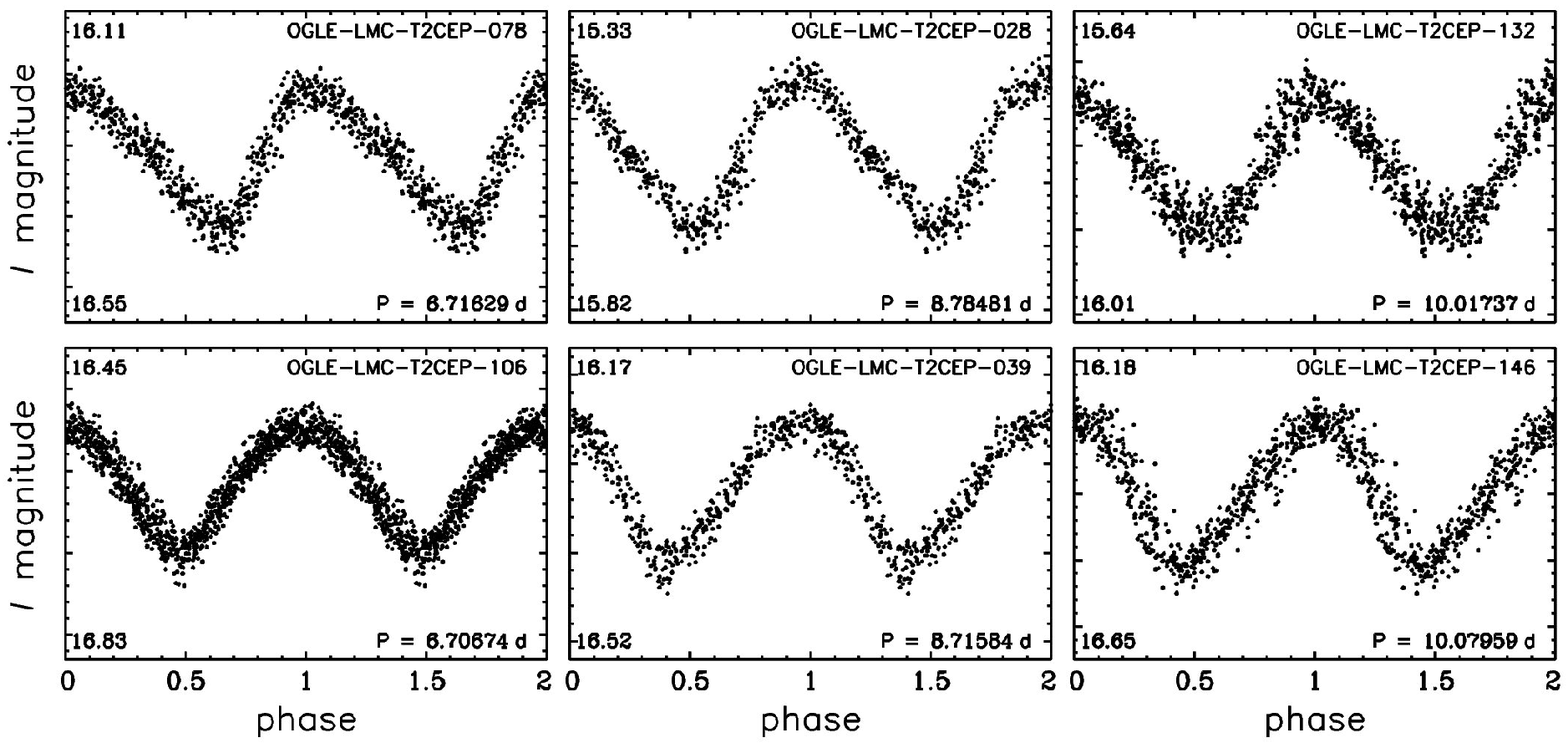}}
\FigCap{Three pairs of light curves of peculiar ({\it upper panels}) and
ordinary ({\it lower panels}) W~Vir stars. Each column presents variables
of similar periods.}
\end{figure}

We separated the peculiar W~Vir stars from the typical ones and noticed
other differences. The different appearance of the light curves shows up in
the diagrams presenting the Fourier parameters \vs periods, especially in
the phase differences $\phi_{21}$ and $\phi_{31}$. We found that using
these parameters it is possible to distinguish two groups of W~Vir stars
also for shorter-period stars, down to 4 days. In total we selected 16
peculiar W~Vir stars with periods in the range 4--10 days. Four of these
stars exhibit additional eclipsing variations superimposed on the pulsation
light curve (see Section~5). Four other peculiar W~Vir stars show secondary
periods in the range of 40--77~days that may be ellipsoidal modulation (in
such a case the orbital periods should be twice as long as the secondary
periods). Taking into account that there must be unfavorable orbital
inclinations for detecting binarity, it is safe to assume that all peculiar
W~Vir stars are members of binary systems. In such a case their current
pulsations could be a consequence of the previous evolution in binary
systems.

In Figs.~1 and 2 the peculiar W~Vir stars are represented by cyan
symbols. As one can notice, these stars are systematically brighter by
about 0.5~mag in the {\it I}-band and 0.7~mag in the {\it V}-band than
typical W~Vir stars of the same periods. Both groups also differ
significantly in $(V-I)$ colors. The peculiar variables tend to be bluer,
although their color distribution seems to be more scattered than for
regular W~Vir stars. The median $(V-I)$ value for the peculiar W~Vir stars
is equal to 0.72~mag, \ie similar to BL~Her stars. There is only little
difference between both groups of W~Vir stars in the $\log{P}{-}W_I$
diagram, because the differences in magnitudes and colors cancel out
in the $W_I$ index.

\Section{``Eclipsing'' Type~II Cepheids}
In our samples of type~II and anomalous Cepheids we failed to identify any
distinct double-mode pulsator with two radial modes simultaneously
excited. Nevertheless, we detected a number of objects with low-amplitude
secondary periodicities. For example, there is an interesting group of
several type~II Cepheids with secondary periods 5--10 longer than pulsation
period. Four of these stars are the peculiar W~Vir stars described in the
previous section. We provide the information about the most prominent
additional periods in the remarks of the catalog.

\begin{figure}[t]
\vspace{-2mm}
\centerline{\includegraphics[width=13.3cm]{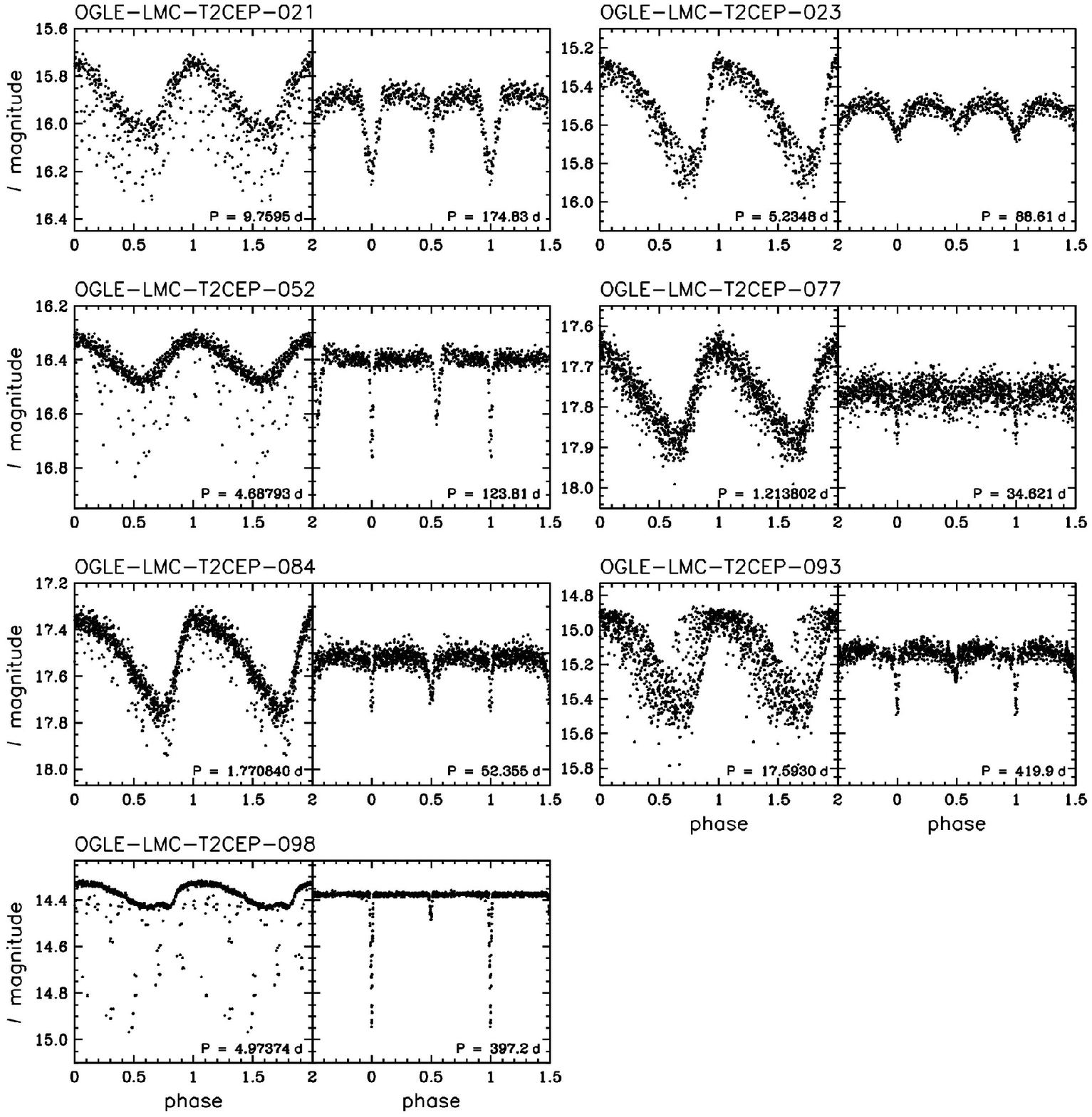}}
\vspace{-2mm}
\FigCap{Light curves of type~II Cepheids with additional eclipsing
variability. In each pair of diagrams {\it left panels} show the original
photometric data folded with the Cepheid periods, while {\it right panels}
show eclipsing light curves after subtracting the Cepheid component.}
\end{figure}

During the search for double-periodic stars we detected seven type~II
Cepheids with eclipsing modulation overimposed on the pulsation
variations. Three of these objects (OGLE-LMC-T2CEP-052, OGLE-LMC-T2CEP-093,
OGLE-LMC-T2CEP-098) were previously reported by Welch \etal (1999). The
light curves of all eclipsing type~II Cepheids are shown in Fig.~6. Here we
present the original {\it I}-band photometry folded with the pulsation
periods and the data after subtraction of the Cepheid light curves and
folded with the orbital periods. The scatter visible in some light curves
is usually caused by variable pulsation periods, and the full solution
requires fitting these changes before subtracting the light curves. We did
it only for the star with the most unstable light curve --
OGLE-LMC-T2CEP-093 -- for which we measured periods and fitted the light
curves separately for each observing season.

It is striking that among 3361 classical Cepheids in the LMC (Paper~I) we
found only three stars with eclipses and one eclipsing system of two
classical Cepheids, while the sample of 197 type~II Cepheids contains as
many as seven eclipsing variables. Of course, confirmation that these
objects are binary systems with a Cepheid as one of the components requires
further studies.

It is even more surprising that four of our ``eclipsing'' type~II Cepheids
have been classified by their Fourier parameters as the peculiar W~Vir
stars described in the previous section. Only three eclipsing/pulsating
stars are regular type~II Cepheids. It is also worth noting that three
Cepheids with eclipsing modulation pulsate with periods between 4 and 6
days, \ie in the frequency minimum between BL~Her and W~Vir stars.

\Section{The Catalog}
\vspace*{12pt}
The catalog data are available through the WWW interface or by anonymous
FTP sites:

\begin{center}
{\it http://ogle.astrouw.edu.pl/} \\
{\it ftp://ftp.astrouw.edu.pl/ogle/ogle3/OIII-CVS/lmc/t2cep/}\\
{\it ftp://ftp.astrouw.edu.pl/ogle/ogle3/OIII-CVS/lmc/acep/}
\end{center}

In the FTP type II Cepheids and anomalous Cepheids are placed in two
separate directories. The variables are arranged according to increasing
right ascension and designated with symbols OGLE-LMC-T2CEP-NNN and
OGLE-LMC-ACEP-NNN (where NNN is a consecutive number) for type~II Cepheids
and anomalous Cepheids, respectively. The files {\sf ident.dat} in each
directory contain coordinates and cross-identifications of the stars with
other catalogs. The following columns give the object designation, OGLE-III
field and the database number of a star, classification (BL~Her, W~Vir or
RV~Tau for type~II Cepheids, F or 1O for anomalous Cepheids), RA and DEC
coordinates for the epoch 2000.0, cross-identification with the OGLE-II
Catalog of Cepheids (Udalski \etal 1999), the MACHO sample of type II
Cepheids (Alcock \etal 1998) and the GCVS vol.~V (Artyukhina \etal
1995). In the last column there are other designations taken from the GCVS.

\begin{figure}[t]
\centerline{\includegraphics[width=12.8cm]{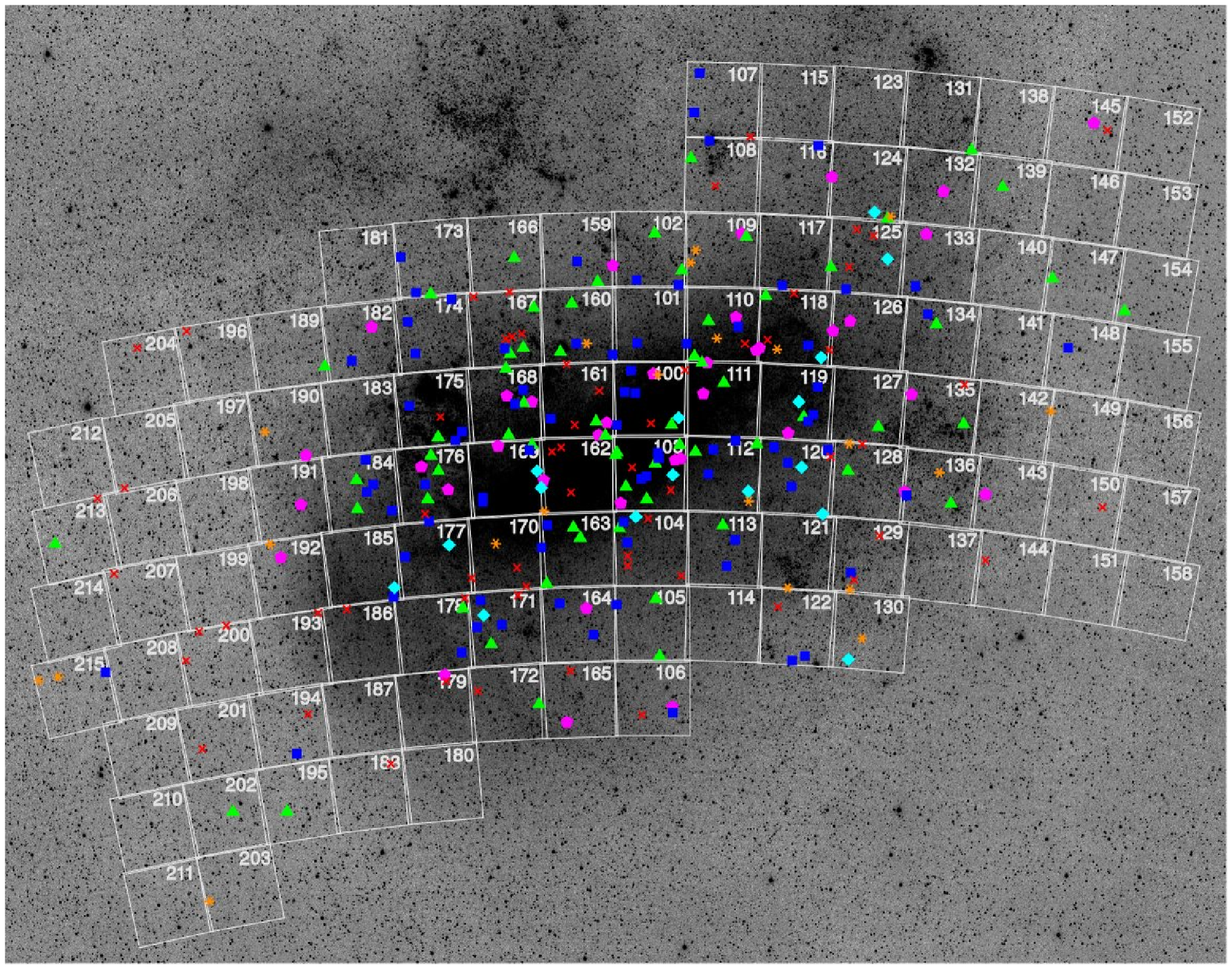}}
\vspace{5mm}
\FigCap{Spatial distribution of type II and anomalous Cepheids in the
LMC. Color symbols represent the same types of stars as in Fig.~1. The
background image of the LMC is originated from the ASAS wide field sky
survey.}
\end{figure}

The parameters of the Cepheids are given in the files {\sf t2cep.dat}, {\sf
acepF.dat} and {\sf acep1O.dat}. The consecutive columns contain: Cepheid
designations, $I$ and {\it V}-band intensity mean magnitudes, periods in
days and their uncertainties, epochs of maximum light, amplitudes in the
{\it I}-band, and Fourier parameters $R_{21}$, $\phi_{21}$, $R_{31}$,
$\phi_{31}$ derived for the {\it I}-band light curves. Periods and their
uncertainties were derived using program {\sc Tatry} by
Schwarzenberg-Czerny (1996). Information about individual objects of
particular interest are provided in the files {\sf remarks.txt}. The
subdirectories {\sf phot/} contain multi-epoch OGLE-II (if available) and
OGLE-III photometry of the stars in our catalog. The subdirectories {\sf
fcharts/} contain finding charts of all objects -- the
$60\arcs\times60\arcs$ subframes of the {\it I}-band DIA reference images,
oriented with N up, and E to the left.

Fig.~7 shows the position of type~II and anomalous Cepheids in the LMC
overplotted on the picture taken by the ASAS survey (Pojma{\'n}ski 1997).
A cross-check of our variables with previously published list of type~II
Cepheids in the LMC revealed 72 objects in common -- 125 Cepheids are
identified for the first time. Our catalog includes all objects detected by
Alcock \etal (1998) in the LMC. The GCVS lists 23 objects in the LMC of
types CWA, CWB, RV or RVA. Our catalog contains 16 of them. Six stars are
outside the OGLE-III fields and one (HV~12904) was reclassified as a
classical Cepheid. On the other hand, object HV~12509, classified in the
GCVS as DCEP (classical Cepheid), occurred to be an ``eclipsing'' (and thus
blended) type~II Cepheid.

It seems that anomalous Cepheids in the LMC were not known to date. As we
mentioned before, the four candidates for stars of that type detected by Di
Fabrizio \etal (2005) are, in our opinion, RR Lyr stars. In the GCVS we
identified only one anomalous Cepheid from our sample, OGLE-LMC-ACEP-009 =
LMC~V0464 = NGC~1786~V1, but it is classified as RRab star, though with a
remark ``BLBOO type? A star of the Galaxy?'' (BLBOO is a designation of
anomalous Cepheids in the GCVS). Nemec \etal (1994), besides NGC~1786~V1,
proposed the variable star NGC~1786~V8 ($P=0.36391$~days) as a candidate
for anomalous Cepheid in the LMC. We identified this object in our
database, and found that 0.36391~days is an evident one-day alias of the
real period: $P=0.57183$~days. NGC~1786~V8 is undoubtedly an RRab star. The
OGLE-II catalog of Cepheids in the LMC (Udalski \etal 1999) contains
several objects from our sample of anomalous Cepheids, but they are
categorized together with type~II Cepheids as ``fainter'' than classical
Cepheids. Thus, in this paper we present the first reliable identifications
of anomalous Cepheids in the LMC. It is worth mentioning that the size of
our sample is comparable with the number of all known anomalous Cepheids in
all the remaining environments.

\Section{Fourier Analysis}
The progression of light curve shapes with periods, clearly visible in
Fig.~3 for type~II Cepheids, is analogous to the Hertzsprung progression
observed for classical Cepheids. Stobie (1973) was the first who noticed
the changes of bump phases with period in the type~II Cepheids. Kubiak and
Udalski (2003) noticed that the progression of light curve shapes of
classical Cepheids, observed in the range of periods between 6 and 24~days,
is reproduced in type~II Cepheids but in the period range 0.9--3~days. The
theoretical models show that such a behavior is caused by the progressive
shift through the 2:1 resonance between the fundamental and second overtone
modes (\eg King \etal 1981, Carson \etal 1981, Carson and Stothers 1982,
Hodson \etal 1982).

The quantitative description of the Hertzsprung progression is possible
thanks to the parameters of the Fourier light curve decompositions (Simon
and Lee 1981). Fourier analysis of type~II Cepheids was performed by
Petersen and Diethelm (1986) and Simon (1986). The standard parameters used
in the light curve analysis are amplitude ratios $R_{k1}=A_k/A_1$ and phase
differences $\phi_{k1}=\phi_k-k\phi_1$, where $A_k$ and $\phi_k$ are
parameters of the truncated Fourier series fitted to the photometric data.

We have fitted Fourier series to the light curves using program {\sc J-23}
by T.~Mizerski. To avoid overfitting the maximum harmonic was adjusted to
minimize the value of $\chi^2$ per degree of freedom. For light curves with
insignificant higher harmonics the amplitude ratios are equal to zero,
while the appropriate phase differences are not defined.

In Figs.~8 and 9 we plot the Fourier parameters $R_{21}$, $\phi_{21}$,
$R_{31}$, $\phi_{31}$ against $\log{P}$. For comparison we include the same
coefficients of fundamental-mode and first-overtone classical Cepheids from
Paper~I. As one can notice, various types of Cepheids follow generally
different sequences in these planes which reflects the progression of light
curves shapes with periods.

For BL~Her stars the $\phi_{21}$ parameter increases slowly with periods in
agreement with the theoretical predictions (Moskalik and Buchler 1993),
while the $\phi_{31}$ grows very fast and reaches value $2\pi$ for
$P\approx1.5$~days. This is the progression center (Petersen and Diethelm
1986) where the 2:1 resonance between fundamental and second overtone modes
is expected. For longer periods $\phi_{31}$ starts growing from zero,
because the phase differences are defined modulo $2\pi$.

\begin{figure}[p]
\centerline{\includegraphics[width=13.5cm, bb=35 55 545 745]{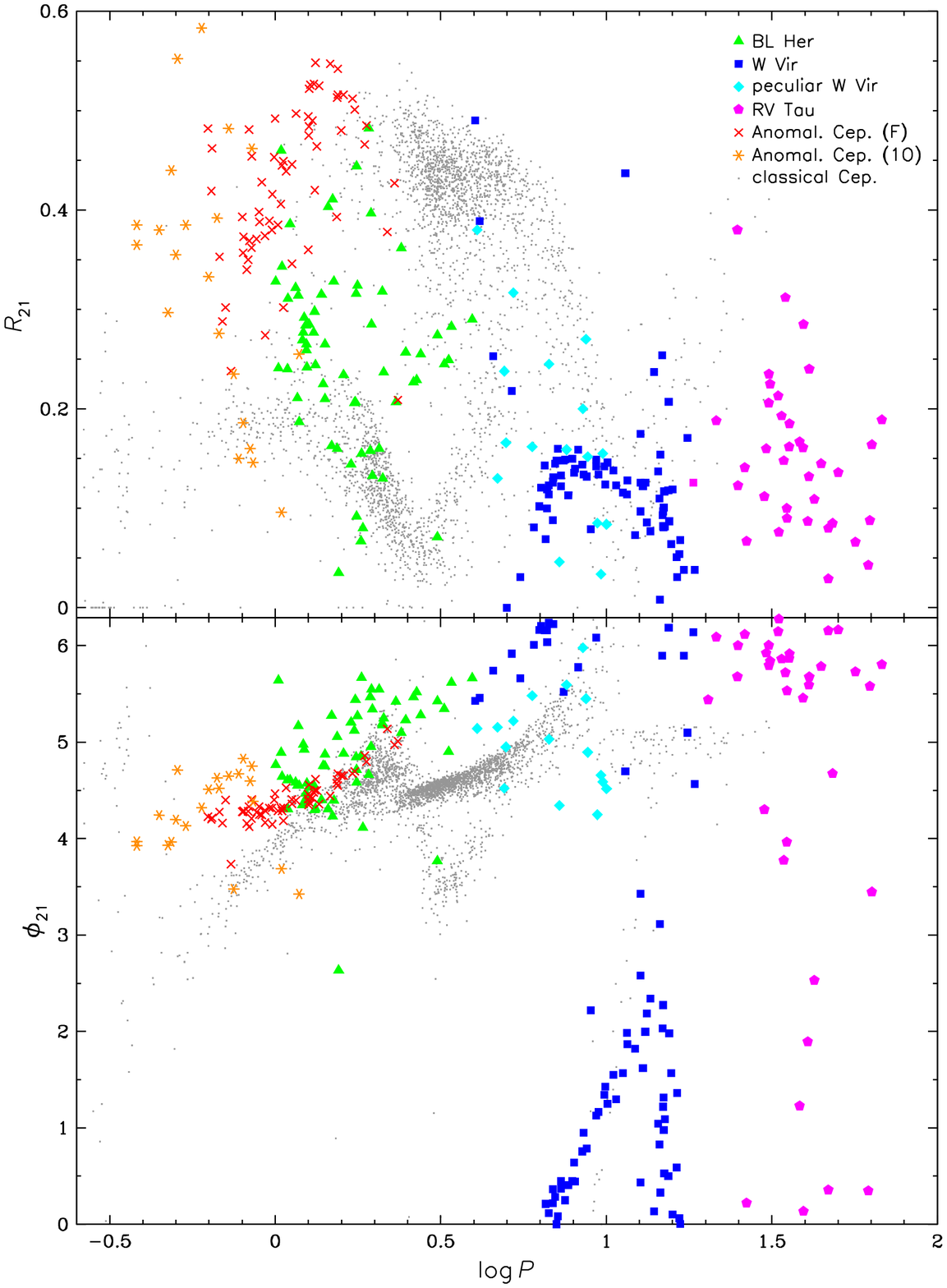}}
\FigCap{Fourier parameters $R_{21}$ and $\phi_{21}$ \vs $\log{P}$ for
Cepheids in the LMC. Symbols are the same as in Fig.~1.}
\end{figure}

\begin{figure}[p]
\centerline{\includegraphics[width=13.5cm, bb=35 55 545 745]{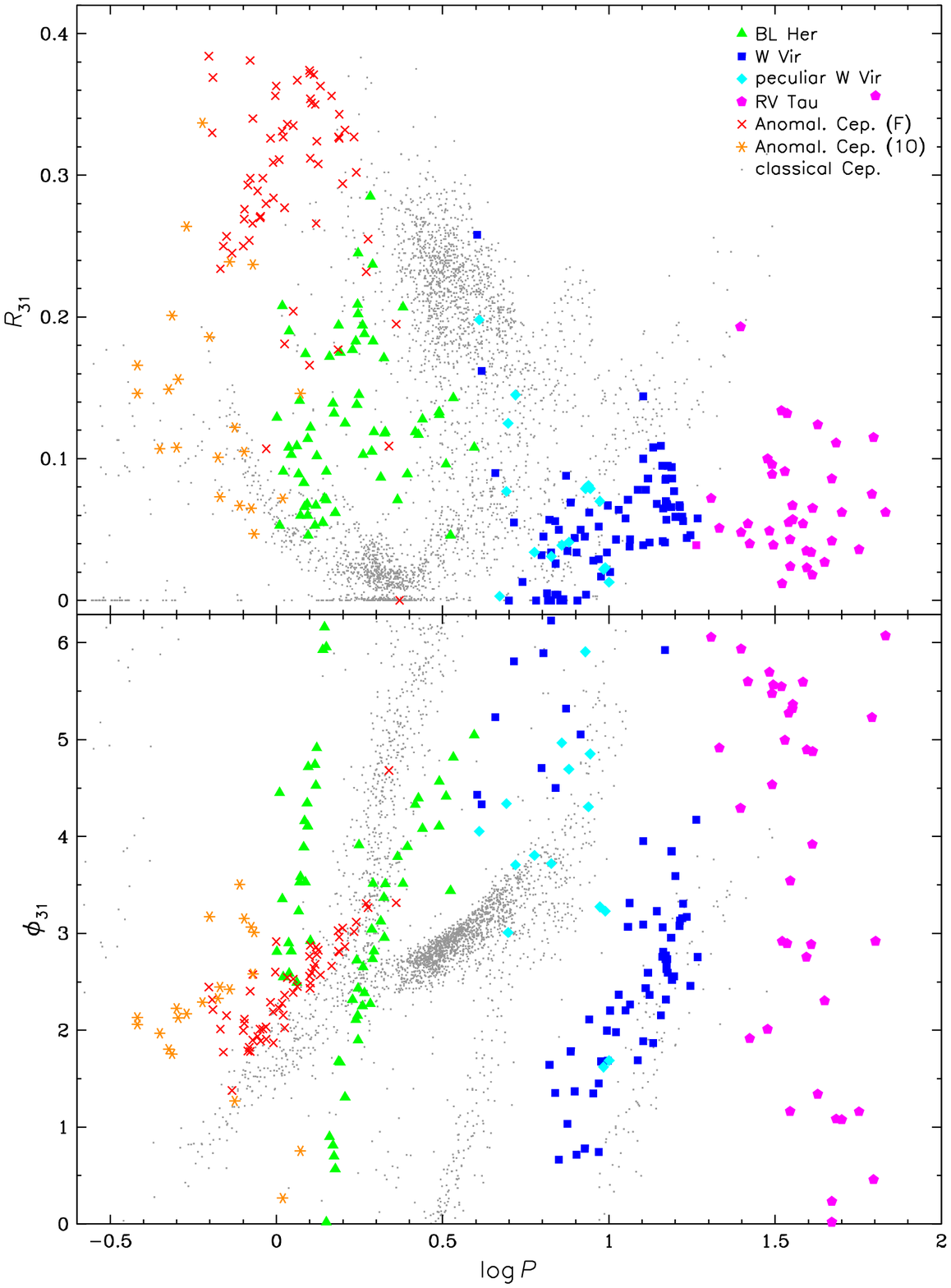}}
\FigCap{Fourier parameters $R_{31}$ and $\phi_{31}$ \vs $\log{P}$ for
Cepheids in the LMC. Symbols are the same as in Fig.~1.}
\end{figure}

The next similar feature occurs for W~Vir stars with periods of about
6~days. Amplitude ratios, as well as amplitudes of the light curves, reach
minimum values for this period, while $\phi_{21}$ and $\phi_{31}$ cross
zero. Similar behavior seem to appear again for W~Vir stars with periods
16--20~days. This time it is better visible in the $\log{P}$--$R_{21}$ and
$\log{P}$--$\phi_{21}$ planes. This latter feature can be interpreted as a
signature of the 2:1 resonance between the fundamental mode and the first
overtone, which is expected to occur at $P\approx17$ (Carson \etal
1981). For RV~Tau stars Fourier parameters are spread over larger area, but
it should be remembered that the uncertainties of these quantities may be
significant due to large scatter of the light curves and alternation of
cycles (the Fourier series was fitted to light curves phased with
``single'' periods).

\Section{Discrimination between Type I and Type~II Cepheids}
\vspace*{-5pt}
The problem of distinguishing between classical and type~II Cepheids is
particularly important for field variables in the Galaxy, because in
general it is impossible to place a given Cepheid in the absolute PL
diagram. A number of criteria were used to identify both types of Cepheids
in the Milky Way -- distance from the Galactic plane, spectral features and
appearance (morphology and stability) of the light curves -- but none of
these method offered unambiguous and complete classification. The large
OGLE-III samples of Cepheids offer the possibility to improve the
distinction criteria, especially these depending on the light curves
shapes. The tool used for quantitative description of light curve shapes
are Fourier decomposition parameters (Figs. 8 and 9).

The problem of discrimination between type I and type~II Cepheids using
their light curves were discussed, among others, by Kwee (1967), Diethelm
(1983), Fernie and Ehlers (1999) and Schmidt \etal (2004). Fernie and
Ehlers (1999) showed that the phase differences $\phi_{i1}$ are good,
although not perfect, discriminants between both types of variables, while
the $R_{i1}$ parameters are useless for this purpose. Schmidt \etal (2004)
discussed various criteria -- morphology of the light curves, Fourier
parameters and light-curve stability -- in terms of differentiation between
type I and type~II Cepheids. They concluded that the shapes of the light
curves are quite useful tool for the classification.

In Figs. 8 and 9 each of the groups: fundamental-mode classical Cepheids,
first-overtone classical Cepheids, type~II Cepheids and anomalous Cepheids,
traces different patterns in the Fourier parameters \vs period diagrams,
however the strips cross each other at various ranges of periods. Thus, it
is not possible to draw lines separating all these classes simultaneously
for all periods, but it is possible to delineate regions in the parameter
space occupied by given types of variables.

For example, type~II Cepheids with periods between 2 and 9~days can be
reliably distinguished by their $\phi_{21}$ parameters, while for
shorter-period type~II Cepheid we recommend using the $\phi_{31}$
coefficient. Probably the best way to separate long-period W~Vir stars
would be using $R_{31}$ parameter. Anomalous Cepheids seem to be well
separated from fundamental-mode classical Cepheids in the
$\log{P}$--$\phi_{31}$ diagram, while it is better to use $R_{31}$
parameter to distinguish them from first-overtone type I Cepheids.

\Section{Summary}
In this paper we extend the OIII-CVS with other pulsating stars in the LMC
populating the upper part of the instability strip -- type~II and anomalous
Cepheids. We significantly increase the number of known type~II Cepheids
in the LMC and discovered the first anomalous Cepheids in this galaxy. Such
large samples will make important contributions to our understanding of
these stars -- their evolution, internal structure and pulsation
mechanisms. The PL relations which can be precisely determined from our
data should open new possibilities for using type II and anomalous Cepheids
as distance indicators.

Our preliminary investigation has shown several new features of the type II
Cepheids. BL~Her and W~Vir stars differ significantly in $(V-I)$ colors,
although there is a subset of W~Vir stars which have, on average, similar
colors as BL~Her stars. This class of peculiar W~Vir stars is also
distinguishable by their light curve shape and luminosity. Seven of the
type~II Cepheids in the LMC exhibit additional eclipsing modulation what is
a very large fraction compared to the sample of classical Cepheids. The
distribution of the Fourier parameters of type II Cepheids shows systematic
pattern with three sharp features for periods around 1.5, 6 and
17~days. Such a behavior is interpreted as an effect of resonances between
two radial modes of pulsation.

\Acknow{The authors would like to thank Prof. W.A.~Dziembow\-ski for helpful 
discussions and a critical reading of the manuscript. We thank
Drs. Z.~Ko{\l}aczkowski, T.~Mizerski, G.~Pojmañski,
A.~Schwar\-zenberg-Czerny and J.~Skowron for providing the software which
enabled us to prepare this study.

This work has been supported by the Foundation for Polish Science through
the Homing (Powroty) Program and by MNiSW grants: NN203293533 to IS and
N20303032/4275 to AU.

The massive period searching was performed at the Interdisciplinary Centre
for Mathematical and Computational Modeling of Warsaw University (ICM UW),
project no.~G32-3. We are grateful to Dr.~M.~Cytowski for helping us in
this analysis.}

\end{document}